\documentclass[aps,jcp,twocolumn,groupedaddress]{revtex4-1}
\usepackage{graphicx}
\usepackage{subfigure}
\usepackage{amsmath}
\usepackage{amssymb}
\usepackage{enumerate}
\usepackage{psfrag}
\usepackage{bm}
\usepackage{color}

\newcommand{\overlap}[2]{{\ensuremath \bigl \langle \,  #1 \, \bigr | \,  #2  \, \bigr \rangle }}
\newcommand{\braket}[3]{{\ensuremath \bigl \langle \,  #1 \, \bigr | \, #2 \, \bigl | \,  #3  \, \bigr \rangle }}
\newcommand{\bigbraket}[3]{{\ensuremath \Bigl \langle \,  #1 \, \Bigr | \, #2 \, \Bigl | \,  #3  \, \Bigr \rangle }}
\newcommand{\ket}[1]{{\ensuremath  \bigl | \,  #1  \, \bigr \rangle }}
\newcommand{\smallket}[1]{{\ensuremath  | \,  #1  \,  \rangle }}
\newcommand{\bra}[1]{{\ensuremath \bigl \langle \,  #1 \, \bigr | }}
\newcommand{\smallbra}[1]{{\ensuremath  \langle \,  #1 \, | }}

\newcommand{\tderiv}{\ensuremath \frac \partial{\partial t}}

\begin{document}


\title{Two-photon ionization of Helium studied with the multiconfigurational time-dependent Hartree-Fock method}
 

\author{David Hochstuhl}
\email[]{hochstuhl@theo-physik.uni-kiel.de}

\author{Michael Bonitz}

\affiliation{Institut f\"ur Theoretische Physik und Astrophysik, D-24098 Kiel, Germany}


\date{\today}

\begin{abstract}
The multiconfigurational time-dependent Hartree-Fock method (MCTDHF) is applied for simulations of the two-photon ionization of Helium. We present results for the single- and double ionization from the groundstate for photon energies in the non-sequential regime, and compare them to direct solutions of the Schr\"odinger equation using the time-dependent (full) Configuration Interaction method (TDCI). We find that the single-ionization is accurately reproduced by MCTDHF, whereas the double ionization results correctly capture the main trends of TDCI.
\end{abstract}

\pacs{23.80 Fb, 32.80 Rm, 03.65.Aa}

\maketitle

\section{Introduction}
The photoionization of Helium is among the best studied processes involving correlated systems in external laser fields. Its investigation dates back over 40 years \cite{Byron_1967} and particularly the last two decades have seen a lot of clarifying work on the experimental as well as on the theoretical front. Owing to these efforts the single-photon ionization is nowadays considered as well understood \cite{Lambropoulos_2008}. Several experiments have been performed, e.g. \cite{Fittinghoff_1992,Walker_1994}, theoretical works provided cross sections which show an excellent agreement with the measured data, e.g. \cite{Foumouo_2006}, and also time-resolved theoretical results are becoming available \cite{Bauch_2010}.

In recent time, the two-photon ionization has gained a lot of attention, being the simplest multiphoton process in which correlation effects play a significant role. Despite the large number of theoretical investigations, see e.g. \cite{Pindzola_1998,Laulan_2003,Foumouo_2006,Nikolopoulos_2007a,Horner_2007,Feist_2008}, it is still far from being understood, which is reflected by several contradicting predictions for differential and total cross sections. Only a few experiments have been performed up to now \cite{Hasewaga_2005,Sorokin_2007,Rudenko_2008}, the results of which were not yet able to definitely decide on the origins of the theoretical discrepancies. 

In this work we present new theoretical results on the two-photon ionization at photon energies in the direct regime, which ranges roughly in between $39.5$ eV -- half of the total groundstate energy of Helium -- and $54.4$ eV -- the second ionization potential. In this regime, double ionization can occur only if the electrons share the energy of the two photons via their Coulomb interaction. The sequential double ionization process, in which the electrons are ionized successively, separated at least by the relaxation-time of the atom, is not possible with only two photons.\\
The simulation of Helium in external laser fields is most directly performed through the time-dependent Schr\"odinger equation (TDSE). The theoretical tools for a direct solution of the TDSE are manifold; most of them employ an expansion of the two-particle wavefunction in terms of coupled spherical harmonics and arrive at a set of equations for the radial functions. The latter is solved either on a two-dimensional grid \cite{Parker_2000} or by expansion in a basis set like Sturmian functions \cite{Foumouo_2006}, B-splines \cite{Laulan_2003}, or the discrete variable representation \cite{Feist_2008}. On the other hand, time-independent approaches have been successfully attempted, for instance the R-matrix Floquet method \cite{Feng_2003}, and lowest-order perturbation theory (LOPT) \cite{Nikolopoulos_2001,Horner_2007}. Last but not least, approximate methods have been applied, such as the time-dependent density functional theory (TDDFT) \cite{Carrera_2009, McKenna_2009} or time-dependent Hartree-Fock (TDHF) \cite{Kulander_1987, Maragakis_1997}. They have the advantage of being applicable to larger atoms, but suffer the main drawback that their accuracy is difficult to estimate.

In this work, we apply the multiconfigurational time-dependent Hartree-Fock (\mbox{MCTDHF}) method, which in principle provides a control of the accuracy. As the name suggests, \mbox{MCTDHF} can be considered as an extension of the time-dependent Hartree-Fock scheme, in which the wavefunction is expressed by several configurations instead of a single one. The origin of the method dates back to the 1980's, where time-dependent multiconfigurational schemes were considered for the first time \cite{Dalgaard_1980,Albertsen_1980,McWeeny_1983,Kosloff_1988}. A particularly successful approach is the multiconfigurational time-dependent Hartree (MCTDH) method introduced by Meyer, Manthe, Cederbaum and coworkers \cite{Meyer_1990,Manthe_1992,Beck_2001}.  This method has later been naturally extended to \mbox{MCTDHF}, where the antisymmetry of the wavefunction is explicitly enforced by using Slater determinants instead of Hartree products as basis states \cite{Zanghellini_2003,Kato_2004,Nest_2005}. For a recent overview on MCTDH and MCTDHF, see the book \cite{Meyer_MultiQM}.

The formulation used in the present work resembles the one given by Alon \emph{et al.} \cite{Cederbaum_unified}, who derived equations based on the second quantization formalism, which, as an aside, have also successfully been applied to the treatment of bosonic particles \cite{Cederbaum_boson}. Here, we present a spin-restricted version of the fermionic \mbox{MCTDHF} equations, which is based upon the same principles, but formulated in terms of spatial orbitals rather than spin-orbitals.

In connection with photoionization, \mbox{MCTDHF} has recently been applied to molecular hydrogen \cite{Kato_2008,NguyenDang_2007} resp. to a version thereof with a screened interaction potential \cite{Jordan_2006}. Some works have also investigated models of reduced dimensionality, see e.g. Refs. \cite{Caillat_2005,Hochstuhl_2010,Bonitz_2010} for one-dimensional and Ref. \cite{Sukiasyan_2009} for two-dimensional model systems. However, to the best of our knowledge, \mbox{MCTDHF} has not been applied to a treatment of the three-dimensional Helium atom in external laser fields so far. In this work we perform such an investigation, the focus of which lies mainly on a critical assessment of the capabilities of the method for the description of correlated photoionization processes. Therefore, we compare the \mbox{MCTDHF} results with the time-dependent (full) Configuration Interaction results, which mark the fully correlated and thus best possible results for a specified single-particle basis.

The paper is organized as follows: In section II, we give a detailed overview on the \mbox{MCTDHF} method in general, our chosen discretization schemes and the numerical implementation. In section III, we present the numerical results for the groundstate and for the two-photon ionization of Helium. Thereafter, in section IV, we give a discussion and outlook. Finally, in the appendix we present the derivation of the spin-restricted \mbox{MCTDHF} equations.

\section{Theory}

\subsection{General considerations}
In this work we are concerned with the solution of the time-dependent electronic Schr\"odinger equation
\begin{align}\label{Schroedinger_eq}
i\partial_t \, \ket{\Psi} \ = \hat H(t) \, \ket{\Psi}\,,
\end{align}
with a Hamiltonian (in atomic units) given by
\begin{align}
\label{Hamiltonian} \hat H(t) \ = \ \sum_{k=1}^N \, \Biggl\{ \frac{{\mathbf {\hat p}_k}^2}2 -  \frac{Z}{r_k}  +  \hat V_\text{ext}(t) \Biggr\}  \, + \,  \frac12  \, \sum_{k\neq l}  \frac{1}{|\mathbf r_k-\mathbf r_l|}\,.
\end{align}
It describes the motion of $N$ electrons in the field of an atomic core with charge $Z$ in the static Born-Oppenheimer approximation. The electrons interact with each other via the two-body Coulomb repulsion, and the operator $\hat V_\text{ext}(t)$ incorporates the action of an external electromagnetic field. 
In second quantization the Hamiltonian reads \cite{Helgaker}
\begin{align}
\label{Hamiltonian_2nd} \hat H(t) \ = \ \sum_{pq} h_{pq}(t) \, \hat E_{pq} + \frac 12 \sum_{pqrs} \, g_{pqrs} \, \hat e_{pqrs}\,,
\end{align}
with the single- and double excitation operators
\begin{align}
\label{single_ex} \hat E_{pq}\, &= \, \sum_\sigma a^\dagger_{p\sigma}a_{q\sigma}\,,\\
\label{double_ex} \hat e_{pqrs} \, &= \, \sum_{\sigma \tau} a^\dagger_{p\sigma}a^\dagger_{r\tau}a_{s\tau}a_{q\sigma}\,,
\end{align}
set up as linear combinations of the \emph{spin-orbital} excitation operators $a^{(\dagger)}_{p\sigma}$, which destroy (create) a particle in the \emph{time-dependent} state $\smallket{\phi_{p\sigma}(t)}$. The Hamiltonian (\ref{Hamiltonian_2nd}) implies a \emph{spin-restricted} treatment, in which the orbitals are independent of their respective spin-projection, i.e. $\smallket{\phi_{p\alpha}(t)} = \smallket{\phi_{p\beta}(t)} =: \smallket{\phi_{p}(t)}$. The matrix elements of the one-body Hamiltonian and the two-body Coulomb operator in this basis are time-dependent as well and given by
\begin{align}
\label{one_body_matrix_el} h_{pq}(t) \, &= \, \int d \mathbf r \; \phi^\ast_{p}(\mathbf r) \left\{-\frac 12 \Delta - \frac Zr + \hat V_\text{ext}(t) \right\} \phi_{q}(\mathbf r)\,,\\
\label{two_body_matrix_el} g_{pqrs} \, &= \, \iint d\mathbf r \, d \mathbf {\bar r} \; \phi^\ast_{p}(\mathbf r) \phi_{q}(\mathbf r) \, \frac1{|\mathbf r -\mathbf {\bar r}|} \, \phi^\ast_{r}(\mathbf {\bar r}) \phi_{s}(\bar {\mathbf r})\,,
\end{align}
where we indicate only the explicit time-dependence due to the external field. The interaction $\hat V_\text{ext}(t)$ with the external field $\boldsymbol {\mathcal E}(t)$ is considered in dipole approximation in the length gauge
\begin{align}
\hat V_\text{ext}(t) \ = \ \boldsymbol {\mathcal E}(t) \cdot \mathbf r\,.
\end{align}
Another common choice is the velocity gauge. The performance of the two gauges is well discussed in the literature, see e.g. \cite{Cormier_1996,Muller_1999}, and will not be analyzed here.


\subsection{The MCTDHF equations}
In the MCTDHF method, the wavefunction is expressed by a linear superposition of \emph{time-dependent} Slater determinants,
\begin{align}\label{MCTDHF_expansion}
\ket{\Psi} \, = \, \sum_{\mathbf n} C_{\mathbf n}(t) \, \ket{n_{1\alpha},\, n_{1\beta},\, \cdots,\, n_{M\alpha}, \, n_{M\beta}; \, t}\,,
\end{align}
which are written in occupation number representation with respect to the spin-orbitals $\{ \smallket{\phi_{p\sigma}(t)} \}_{1\leq p\leq M}$, from which they directly inherit their time-dependence. The sum runs over the complete set of $\binom M{N_\alpha} \binom M{N_\beta}$ Slater determinants, that can be constructed from $N_\alpha$ particles with spin-projection $\alpha$, $N_\beta$ particles with spin-projection $\beta$ and $M$ spatial orbitals. Their spin-projection is consequently restricted to $S_z=(N_\alpha-N_\beta)/2$. Alternatively, in Eq.~(\ref{MCTDHF_expansion}) we could also use an expansion in configuration state functions, which are not only eigenfunctions of the spin-projection operator (as Slater determinants are), but also of the total spin operator \cite{Helgaker}.

The MCTDHF ansatz is characterized by two sets of time-dependent variational parameters, namely by the expansion coefficients $C_\mathbf{n}(t)$ and by the orbitals $\smallket{\phi_p(t)}$, which determine the actual many-body basis. Their equations of motion are found with the Lagrange formulation of the time-dependent variational principle which requires a minimization of the action functional \cite{Cederbaum_boson,Cederbaum_unified}
\begin{align}
\notag &S\Bigl[\bigl \{C_{\mathbf n}(t) \bigr \}, \bigl \{\smallket{\phi_p(t)}\bigr\} \Bigr] \ = \ \int dt \ \Bigg \{ \bigbraket{\Psi}{\hat H - i\frac{\partial}{\partial t}}{\Psi} \\
\label{action_func}  &\qquad \qquad -  \, \sum_{kl} \mu_{kl}(t) \, \Bigl( \overlap{\phi_k}{\phi_l} -\delta_{kl} \Bigr) \Bigg \}
\end{align}
with respect to the variational parameters. The Lagrange multipliers $\mu_{kl}(t)$ are introduced to ensure the orthonormality of the orbitals during the temporal evolution. The procedure of taking functional derivatives is performed in detail in Appendix \ref{AppA}. It results in the spin-restricted MCTDHF equations
\begin{gather}
\label{MCTDHF_coefficient} i  \, \dot C_{\mathbf n}(t) \ = \ \sum_{\mathbf m} \braket{\mathbf n}{ \hat H(t) }{\mathbf m} \; C_{\mathbf m}(t)\,, \\
\label{MCTDHF_orbital} i \,  {\ket{\dot \phi_n}} \, = \, \hat{\mathbf  P} \, \Biggl\{ \, \hat h(t) \ket{ \phi_n} +  \sum_{pqrs} \left( {\mathbf D}^{-1}\right)_{np} d_{pqrs} \, \hat g_{rs} \ket{ \phi_q} \Biggr\}\,,
\end{gather}
a set of coupled first-order differential equation for the temporal evolution of the variational parameters. Equation (\ref{MCTDHF_coefficient}), here denoted as wavefunction equation, is just the Schr\"odinger equation represented in the (time-dependent) Slater determinant basis. The nonlinear equation (\ref{MCTDHF_orbital}) will be called the orbital equation.
In it we introduced the one- and two-particle density matrices
\begin{align}
\label{Dens1} D_{pq} \ &= \  \braket{\Psi}{\hat E_{pq}}{\Psi}\,, \\
\label{Dens2} d_{pqrs} \ &= \ \braket{\Psi}{\hat e_{pqrs}}{\Psi}\,,
\end{align}
as well as the mean-field potential operator $\hat g_{rs}$, which in coordinate representation reads
\begin{align}
\label{mean-field} g_{rs}(\mathbf r) \ = \ \int d\bar {\mathbf r} \; \phi^\ast_r(\bar {\mathbf r}) \, \frac{1}{|\mathbf r -\bar {\mathbf r}|} \, \phi_s(\bar {\mathbf r})\,.
\end{align}
Particularly in the literature on (single-determinant) Hartree-Fock, $g_{rs}(\mathbf r)$ is also called Coulomb- ($r=s$) and exchange-integrals ($r\neq s$).
Further, in Eq.~(\ref{MCTDHF_orbital}) the elimination of the Lagrange multipliers led to a projection operator
\begin{align}
\label{projector}\hat{\mathbf P} \ = \ 1 \, - \, \sum_{m=1}^M \, \ket{\phi_m}\bra{\phi_m}\,,
\end{align}
which projects on the orthogonal complement of the subspace spanned by the orbitals.

Before we proceed in the numerical solution of the MCTDHF equations, let us take a comparison of the present method with the (full) Configuration Interaction (CI) method. The latter also employs a linear expansion of the wavefunction, however in a basis of \emph{static} Slater determinants. They are constructed from a time-independent single-particle basis $\{\smallket{\psi_{k\sigma}}\}$ of size $N_b$ (instead of $M$), leading to a size of the determinant basis of $\binom {N_b}{N_\alpha} \binom {N_b}{N_\beta}$. As the simulation of ionization processes typically requires a large single-particle basis in order to provide an adequate description of the scattering states (in this work we use up to $N_b \sim 10000$), the many-body Hilbert space quickly becomes prohibitively large causing CI to be applicable at most to two-particle systems. In contrast, MCTDHF may provide a much more compact representation of the wavefunction by using determinants constructed from $M$ time-dependent orbitals of the form
\begin{align}
\label{basis_expansion} \ket{\phi_n(t)} \ = \ \sum_{j=1}^{N_b} b_{nj}(t) \, \ket{\psi_j}\,.
\end{align}
As $M$ is typically much smaller than $N_b$, the corresponding Slater determinant basis is considerably reduced, while the space that can be accessed by the single-particle basis is practically identical. In this way, it is possible that the orbitals match the physical intuition: an orbital could for example be composed dominantly of a bound state and, at the same time, also contain an ionized fraction, as it could be induced by an external field. Orbitals adapted in such a way are then likely to yield a more compact and appropriate Slater determinant basis, so that the wavefunction can hopefully be expressed adequately by a small set of optimized determinants and not by a large static basis. Lastly, the efficiency of the MCTDHF approach depends on the degree of correlation present in the wavefunction.

\subsection{Discretization of the orbital equation}
Once the orbitals are discretized through the expansion  in an orthonormal basis (\ref{basis_expansion}), upon insertion into the orbital equation and projection onto $\smallbra{\psi_k}$ we readily obtain an equation governing the time-dependence of the coefficients,
\begin{align}
\notag i \,\dot b_{nk}(t) \, &= \, h^{(1)}_{nk}(t) - h^{(3)}_{nk}(t) \ + \\
&\qquad \, \sum_{pqrs} \, \left(\mathbf D^{-1}\right)_{np} d_{pqrs} \; \Bigl( g^{(3)}_{kqrs}-g^{ (5)}_{kqrs} \Bigr)\,.
\end{align}
The superscripts label the number of performed transformation steps. Let the electron integrals in the time-independent basis $\{\smallket{\psi_k}\}$ be denoted by $\mathfrak{h}$ and $\mathfrak{g}$, respectively, and defined analogously to Eqs.~(\ref{one_body_matrix_el}) and (\ref{two_body_matrix_el}). The one-body matrix elements are then given by
\begin{align}
\label{trafo_h1} h^{(1)}_{nk}(t) \ &= \  \sum_{j=1}^{N_b}  b_{nj} \ \mathfrak{h}_{jk}(t)\,, \\
\label{trafo_h2} h_{mn}(t) \ &= \  \sum_{k=1}^{N_b}  b^\ast_{mk} \ {h}^{(1)}_{nk}(t)\,, \\
h^{(3)}_{nk}(t) \ &= \  \sum_{m=1}^{M}  b_{mk} \ {h}_{mn}(t)\,,
\end{align}
while the two-body quantities read
\begin{align}
\label{g3_1} g^{(3)}_{iqrs} \  &= \ \int d\mathbf r \; \psi^\ast_i(\mathbf r) \, \phi_q(\mathbf r) \; g_{rs}(\mathbf r)\\
\label{g3_2} &= \  \sum_{j,k,l=1}^{N_b}  b_{qj} \, b^\ast_{rk} \, b_{sl}  \ \mathfrak{g}_{ijkl}\,,\\
\label{trafo_g4} g_{pqrs} \ &= \   \sum_{i=1}^{N_b} b^\ast_{pi} \;  g^{(3)}_{iqrs}\,,\\
\label{trafo_g5}  g^{(5)}_{iqrs} \ &= \  \sum_{p=1}^{M}  b_{pi} \;  g_{pqrs}\,.
\end{align}
The quantities without superscripts, Eq.~(\ref{trafo_h2}) and (\ref{trafo_g4}), are the one- and two-electron integrals in the time-dependent basis already encountered in the Hamiltonian~(\ref{Hamiltonian_2nd}). The transformations (\ref{trafo_h1})-(\ref{trafo_g5}) need to be performed at each propagation step and constitute the main time-consuming part with respect to the single-particle basis. In particular, the major effort is the calculation of the mean-field potential $g_{rs}(\mathbf r)$, Eq.~(\ref{mean-field}), which corresponds to evaluation of the first two summations in Eq.~(\ref{g3_2}). It grows as $\mathcal O(N_b^{\,4})$ for a general basis, like Gaussian- or Slater type orbitals, Sturmians, etc., and even though there exist techniques to reduce the effort by one order to $\mathcal O(N_b^{\,3})$ -- for instance by using a Cholesky decomposition of the two-electron matrix \cite{Wilson_1990} -- we found that the size of the basis is still limited to, say, $N_b \sim 150$, which is too small for our purpose.

\begin{figure}[!t]
  \begin{center}
    \includegraphics[width=0.48\textwidth]{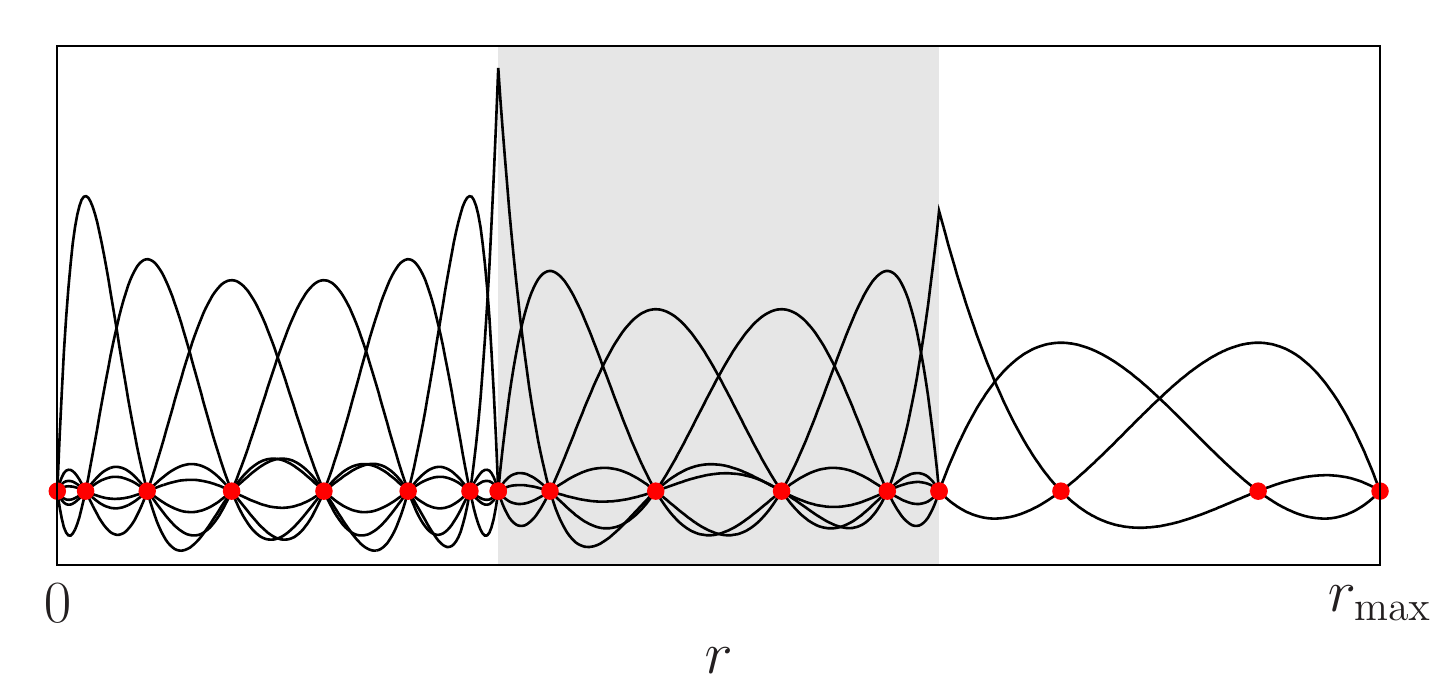}
  \end{center}
\caption{\label{fig:FEDVR}(Color online) Illustration of the finite-element discrete variable representation basis used to discretize the radial part of the wave function. Shown are three equally spaced finite elements with $8, 6, 4$ Gauss-Lobatto points, respectively, (red) dots. Note that the basisfunctions are zero at all gridpoints except one, and that the functions corresponding to the first and last gridpoint have been removed in order to satisfy the boundary conditions.}
\end{figure} 

\subsection{DVR product basis}
In the following, we show how to overcome the afore mentioned problems by using an appropriate single-particle basis. For the treatment of three-dimensional atoms, a spherical expansion provides a description that can be much more compact than in cartesian coordinates \cite{Muller_1999}, owing to the fact that the wavefunction is often very smooth perpendicular to the radial direction. Hence, we use an expansion in spherical harmonics $Y_{l m}$,
\begin{align}\label{3D_basis_expansion}
\phi_p(\mathbf r,t) \ = \ \sum_{kl m} \ b_{p,kl m}(t) \, \frac{\chi_k(r)}{r} \; Y_{l m}(\Omega)\,.
\end{align}
The radial coordinate is described by a discrete variable representation (DVR) basis $\chi_k(r)$. As the DVR is in common use for roughly 30 years in quantum mechanics \cite{Lill_1982,Light_1985} and was recently successfully used also in many-body calculations \cite{Balzer_2010a,Balzer_2010b}, we refer the reader to the cited publications for a general introduction. We only note its main advantage, namely the diagonal representation of spatially local operators,
\begin{align}
\braket{\chi_m}{f(r)}{\chi_n} \ = \ \delta_{mn} \; f(r_m) \,,
\end{align}
where the diagonal entries are simply given by the function values on a DVR grid. The DVR thus resembles the representation on a finite difference grid, but with an important advantage: while finite difference schemes converge polynomially with the number of gridpoints, the DVR may provide an exponential convergence behaviour. This is related to the representation of nonlocal operators, like derivative operators, which in general are given by full matrices.

To further take advantage of this feature, but avoid full matrices, we particularly use the finite element DVR (FEDVR) of Rescigno \emph{et al.} \cite{Rescigno_2000,McCurdy_2004,Schneider_2006}. The corresponding grid of size $N_{rad}$ consists of finite elements of Gauss-Lobatto nodes, see Fig.~\ref{fig:FEDVR}. The advantage in using finite elements is that the kinetic energy matrix attains a block structure, as matrix elements of one-particle operators are nonzero only within adjacent elements. As it is shown in \cite{Feist_dissertation}, by choosing a reasonable distribution and size of the finite elements one can avoid large energy eigenvalues of the single-particle Hamiltonian, which would make the MCTDHF equations more stiff and slow down the time-evolution. Further, the positioning of the finite elements can be matched to the physical problem, e.g. by employing a denser grid near the atom core.

Having introduced the product basis, the main idea to improve the efficiency is now to employ the Poisson equation for the calculation of the mean-field potential, as it is done routinely in (single-determinant) Hartree-Fock implementations, see e.g. \cite{Kulander_1987,Shiozaki_2007}. Upon application of the Laplace operator to Eq.~(\ref{mean-field}), we readily obtain
\begin{align}\label{Poisson}
\Delta g_{rs}(\mathbf r) \ = \ -4\pi \; \phi^\ast_r(\mathbf r)\, \phi_s(\mathbf r)\,.
\end{align}
We solve this equation by expansion of $g_{rs}(\mathbf r)$ in the basis (\ref{3D_basis_expansion}). Note that here the grid-like DVR treatment is essential, as it provides sufficient flexibility in representing the mean-field potential. The discretized version is treated with a method similar to the one used in Ref.~\cite{Becke_1988}. It involves the solution of linear systems having a banded structure, which is caused by the finite element basis. Thereafter, the integral (\ref{g3_1}) is calculated using the DVR quadrature rule and analytical formulas for the product of three spherical harmonics \cite{Arfken_Weber}. All in all, we obtain a scaling of $\mathcal O(N_{rad} \, N_{ang})$, [here $ N_{ang}$ denotes the number of angular basis functions], that is linear with respect to $N_{rad}$, thus allowing for the application of rather large radial grids.
We finally note, that particularly for the chosen FEDVR + spherical harmonic basis, McCurdy \emph{et al.} have derived a similar method for the calculation of the Coulomb repulsion integrals~\cite{McCurdy_2004}, which may also be exploited in the solution of Eq.~(\ref{Poisson}).

\subsection{Numerical implementation}
We briefly mention the basic ideas and properties of our program \emph{Kiel MCTDHF}. It is able to treat both fermions and bosons in a unified way where, for the latter, an expansion in permanents instead of Slater determinants is used. For fermions, it provides the spin-restricted treatment employed in this work, as well as the spin-unrestricted version of the MCTDHF equations. Beside the single-particle basis set used here, several others such as Gaussian- and Slater type orbitals, a spheroidal DVR grid and some one-dimensional DVRs have been implemented, which allow for a treatment of \mbox{one-,} \mbox{two-,} and three-dimensional systems. The structure of the MCTDHF equations thereby allows for an implementation of the different basis sets independent of the wavefunction part, i.e. of the solution of Eq.~(\ref{MCTDHF_coefficient}). The time-propagation of the equations of motion is performed with the general purpose integrators of Ref.~\cite{Numerical_recipes}, of which we apply most of the times an eight-order Runge-Kutta method with adaptive stepsizes. The initial groundstate is found via propagation in imaginary time starting from a random state.

For the evaluation of the rhs. of the wavefunction equation (\ref{MCTDHF_coefficient}), we mainly follow the Configuration Interaction techniques described in Ref.~\cite{Helgaker}: the Hamiltonian matrix is calculated using the minimal operation count (MOC) method, which utilizes a separation of the Slater determinants in products of $\alpha$- and $\beta$-spin strings. Due to the absence of spin-interactions in the Hamiltonian~(\ref{Hamiltonian}), we use only the configurations with vanishing spin-projection $S_z=0$ in the expansion of the wavefunction~(\ref{MCTDHF_expansion}).

As the Hamiltonian and the density matrices have to be constructed in each time-step, the matrix elements of the excitation operators, Eqs.~(\ref{single_ex}) and (\ref{double_ex}), are required repeatedly. They are determined at the beginning of the run by using a graphical representation method and stored in the following. Due to the representation of the determinants in spin-strings, only little memory is needed for their storage.

The solution of the orbital equation is found as discussed above. Its effort scales as $\mathcal O(M^2 N_b N_\text{ang})$, what is much larger than the size of the determinant basis given by $\binom M1 \binom M1 = M^2$. The overall performance of the calculations presented here is thus determined by the orbital equation. For larger atoms, however, this situation is inverted due to the exponential scaling of the many-body basis size.

\subsection{Time-dependent Configuration Interaction}
As noted above, the helium atom has been subject of many investigations based on a direct solution of the Schr\"odinger equation and hence, there are many results available which could serve as a reference. Nevertheless, we implemented our own direct solution method, in order to apply the same single-particle basis as for the MCTDHF calculations and to focus exclusively on the effects of the many-body basis. This way, we achieve a more consistent comparison, which is particularly independent of the single-particle basis and which can be performed strictly on a numerical level.

The actual solution of the Schr\"odinger equation is found using the time-dependent (full) Configuration Interaction (TDCI) method, which has been briefly discussed above. Our approach offers the appealing advantage, that the basic routines can be adopted from MCTDHF without any changes. The time-independent eigenvalue equation is solved by a restarted Lanczos algorithm with full re-orthogonalization \cite{Saad_1992}. Subsequently, the groundstate is propagated using the short-iterative Lanczos method, which basically employs the same algorithm to approximate the exponential time-evolution operator \cite{Park_1986}. In a spherical harmonic basis like (\ref{3D_basis_expansion}), the method becomes equivalent to a time-dependent close-coupling (TDCC) treatment, the only difference is that the angular part is represented by the uncoupled quantum numbers $(l_1,l_2,m_1,m_2)$ instead of the coupled set $(L,M,l_1,l_2)$.

\section{Numerical results}


\subsection{Extraction of physical quantities}
The route towards observables is, in principle, straightforward:  given the final wavefunction, all information can be obtained by projection on subsets of the complete spectrum of atomic eigenstates. For example, single and double ionization events can be identified by projecting on the corresponding singly and doubly ionized Helium eigenstates. The determination of these states, however, would require the full diagonalization of the Hamiltonian, which is far beyond reach.
One must, therefore, resort to a projection on approximate states; these include uncorrelated eigenstates, e.g. antisymmetrized products of single-particle eigenfunctions \cite{Colgan_2004,Feist_2008} as well as, less frequently, states that contain a certain amount of correlation~\cite{Laulan_2003,Foumouo_2006}. It is important to note that the results from both approaches differ significantly, and that the discussion on the correct treatment has not settled yet \cite{Foumouo_2006,Feist_2008,Lambropoulos_2008}. As we are mainly interested in the capabilities of the MCTDHF scheme, in this work, we apply the simpler alternative and neglect correlation in the ionized eigenstates.

In the MCTDHF method, rather than employing the many-body wavefunction, it is convenient to use the single- and two-particle density matrices which have to be supplied at each propagation step in the solution of the orbital equation (\ref{MCTDHF_orbital}).
With them, the expectation values of all single- and two-particle operators are available. In coordinate representation the corresponding single- and two-particle densities are given by \cite{Helgaker}
\begin{gather}
\label{single_dens} D(\mathbf r,t) \; = \; \sum_{pq} \, D_{pq}(t) \, \phi^\ast_p(\mathbf r,t) \phi_q(\mathbf r,t)\,,\\
\label{double_dens} d(\mathbf r,\mathbf r^\prime,t) \, = \, \frac 12 \, \sum_{pqrs} d_{pqrs}(t) \, \phi^\ast_p(\mathbf r,t) \phi_q(\mathbf r,t) \phi^\ast_r(\mathbf r^\prime,t) \phi_s(\mathbf r^\prime,t) \,.
\end{gather}
Note that for two-particle systems, the two-particle density is equal to the absolute square of the wavefunction $|\Psi(\mathbf r,\mathbf r^\prime,t)|^2$. With the densities, we have access to the single- and double ionization yield, defined as the fraction of the densities outside a chosen radius $R_0$,
\begin{gather}
\label{single_ion}P_1(t) \ = \ \int_{r>R_0}  D(\mathbf r,t) \ d\mathbf r \,,\\
\label{double_ion}P_2(t) \ = \ \int_{r>R_0} \int_{r^\prime>R_0}  d(\mathbf r,\mathbf r^\prime,t) \ d\mathbf r\, d\mathbf r^\prime \,.
\end{gather}
Analogous definitions have been used in Ref.~\cite{Parker_2000}. Similarly, it is also possible to study angular distributions by integrating only over a specified solid angle.
As mentioned above, the formulas (\ref{single_ion},\ref{double_ion}) can also be viewed as resulting from a projection, in which the singly and doubly ionized continuum states are assumed to be localized in different spatial regions. Obviously, this is a crude approximation: for instance, bound Rydberg states, which extend over a large spatial region, may contribute to the ionization yield, and vice versa, delocalized continuum states contribute to the bound fraction. Nevertheless, it has been possible to relate the thus obtained results to experimental data~\cite{Parker_2000}.
It is further advantageous, that Eqs.~(\ref{single_ion},\ref{double_ion}) are directly applicable also to larger atoms, where projection schemes would require a specific set of scattering states.

\begin{table}[!t]
 \begin{center}
  \begin{tabular}{c c c}
	\hline
	\hline
	\parbox[0pt][1.6em][c]{0cm}{} approximation & energy [a.u.] & correlation\\
	\hline
	\parbox[0pt][1.9em][c]{0cm}{} $\phantom{xxx}$ HF $\phantom{xxx}$  & $\phantom{xxx} $ $\ -2.86178 \ $ $\phantom{xxx}$& $\phantom{xxx}$ $\ 0 \, \% \ $ $\phantom{xxx}$\\
	\parbox[0pt][1.6em][c]{0cm}{} $M=2$ & $\ -2.87805 \ $ & $\ 40  \, \% \ $\\
	\parbox[0pt][1.6em][c]{0cm}{} $M=3$ & $\ -2.88484 \ $ & $\ 56 \, \% \ $\\
	\parbox[0pt][1.6em][c]{0cm}{} $M=4$ & $\ -2.89135 \ $ & $\ 72  \, \% \ $\\
	\parbox[0pt][1.6em][c]{0cm}{} $M=5$ & $\ -2.89775\ $ & $\ 87 \, \% \ $\\
	\parbox[0pt][1.6em][c]{0cm}{} $M=7$ & $\ -2.89905\ $ & $\ 91 \, \% \ $\\
	\parbox[0pt][1.6em][c]{0cm}{} $M=9$ & $\ -2.90017\ $ & $\ 93  \, \% \ $\\
	\parbox[0pt][1.6em][c]{0cm}{} $M=15$ & $\ -2.90210\ $ & $\ 98 \, \% \ $\\
	\parbox[0pt][1.6em][c]{0cm}{} CI & $\ -2.90285 \ $ & $\ 100 \, \% \ $\\
	\hline
	\parbox[0pt][1.6em][c]{0cm}{} exact \cite{Sims_2002} & $\ -2.90372 \ $ & \\
	\hline
	\hline
 \end{tabular}
 \end{center}
 \caption{\label{tab:energies}Helium groundstate energies for different numbers $M$ of time-dependent orbitals, and percentage of the covered correlation energy. The difference between our Configuration Interaction result and the exact nonrelativistic groundstate energy is due to the angular basis, which includes partial waves only up to $l=2$.}
\end{table}

\subsection{Groundstate}
In this work, the time-evolution is started from the groundstate, which itself is found by imaginary time propagation (ITP) of a random state according to the MCTDHF equations. As the corresponding time-evolution operator is non-unitary, at each propagation step the wavefunction coefficients have to be renormalized. In addition, we orthonormalize the orbital expansion coefficients, although this is in principle not necessary, owing to the constraint in the action functional~(\ref{action_func}). However, it yields a more stable scheme and prevents small numerical errors from growing during the ITP.

In Tab.~\ref{tab:energies}, we present the groundstate energies for different numbers $M$ of MCTDHF orbitals. In each calculation, we use a single-particle basis including all partial waves up to $l=2$, and a radial part which is described by three finite elements of length $4$ a.u. with $11$ Gauss-Lobatto nodes, that span the region up to $r_\text{max}=12$ a.u. This is a typical gridpoint distribution also employed in the time-dependent calculations presented below, i.e. not optimized for groundstate calculations. The results are compared to the full CI energy, which marks the fully correlated result for the chosen single-particle basis. In the second row of the table, we give the fraction of the correlation energy, defined as the difference between the CI and the HF result, that is covered by the respective approximation.

With increasing numbers of self-consistently determined MCTDHF orbitals, the energies obviously converge against the CI reference result. The first correction to Hartree-Fock, $M=2$, is thereby able to account for $40 \,\%$ of the total correlation energy. On the other hand, the most accurate approximation considered in this work, $M=15$, covers $98 \,\%$ of the correlation energy and differs from the CI energy only by $0.8$ mHa. This meets the (loosely specified) requirements of chemical accuracy, and should be appropriate for most applications.

The observed convergence is substantially slower than for one-dimensional model atoms \cite{Hochstuhl_2010,Bonitz_2010}; we have verified by using different initial states and basis sets that the groundstate has indeed been reached. Next, we study whether the description of time-dependent processes is similarly accurate.

\begin{figure}[!t]
  \begin{center}
    \includegraphics[width=0.48\textwidth]{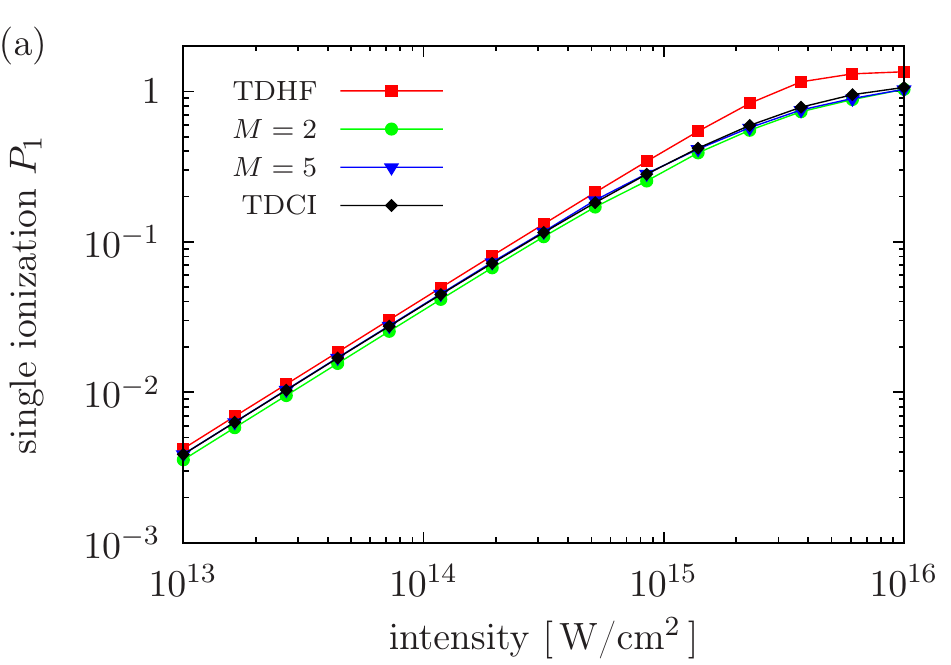}
    \includegraphics[width=0.48\textwidth]{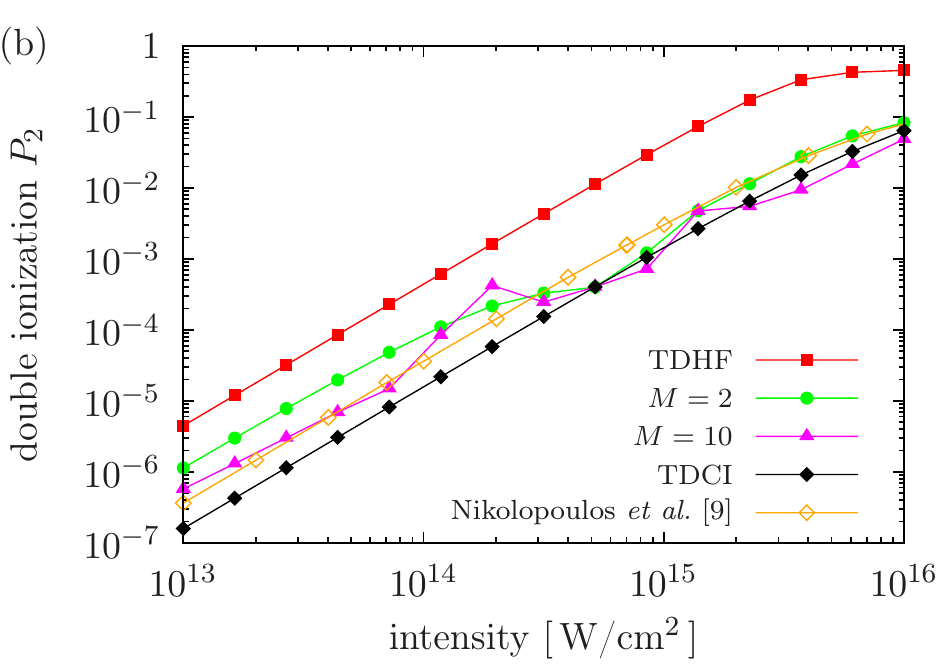}
  \end{center}
\caption{\label{fig:ion_vs_int}(Color online) (a) Single- and (b) double ionization yield vs. intensity for a photon energy of $45$ eV. Shown are the results for different MCTDHF approximations and TDCI as well as the results of Nikolopoulos and Lambropoulos, Ref. \cite{Nikolopoulos_2007a}.}
\end{figure}
 
\begin{figure}[!t]
  \begin{center}
    \includegraphics[width=0.48\textwidth]{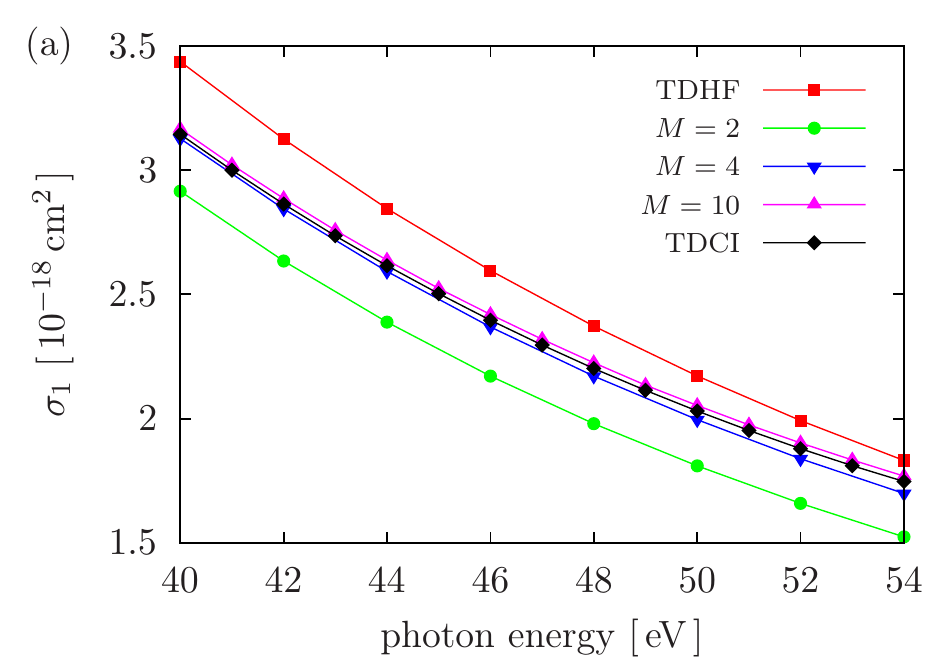}
    \includegraphics[width=0.48\textwidth]{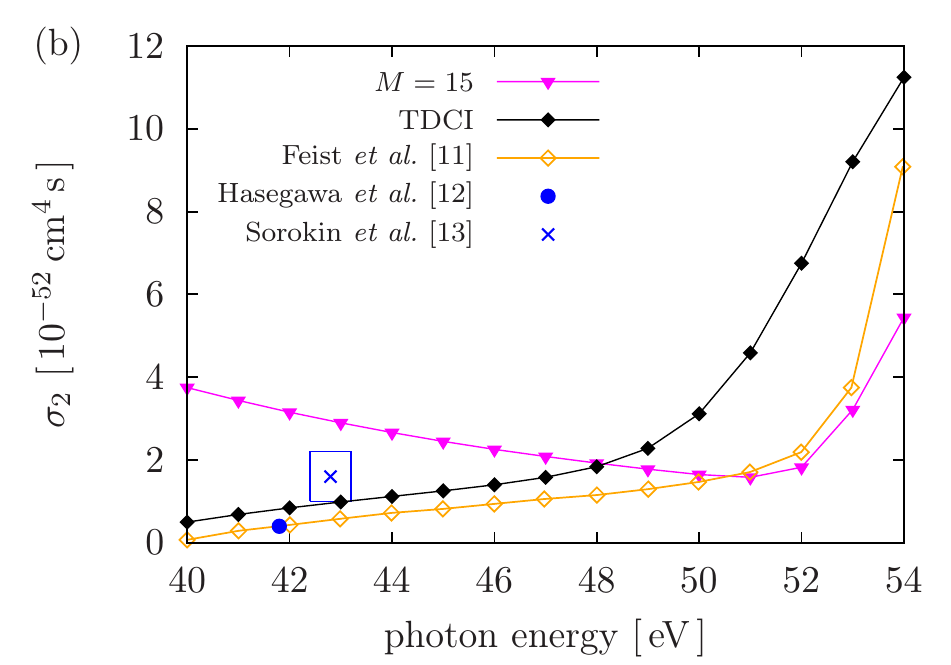}
  \end{center}
\caption{\label{fig:ion_vs_freq}(Color online) (a) Single- and (b) double total ionization cross sections vs.  photon energy, at a fixed field intensity of $10^{13}$ W$/$cm${}^2$. Shown are the results of selected MCTDHF approximations and from TDCI in the same basis, the results of Feist \emph{et al.} \cite{Feist_2008}, as well as the results of experiments by Hasegawa \emph{et al.} \cite{Hasewaga_2005} and Sorokin \emph{et al.} \cite{Sorokin_2007} (the latter with errorbars).}
\end{figure}

\begin{figure}
  \begin{center}
    \includegraphics[width=0.48\textwidth]{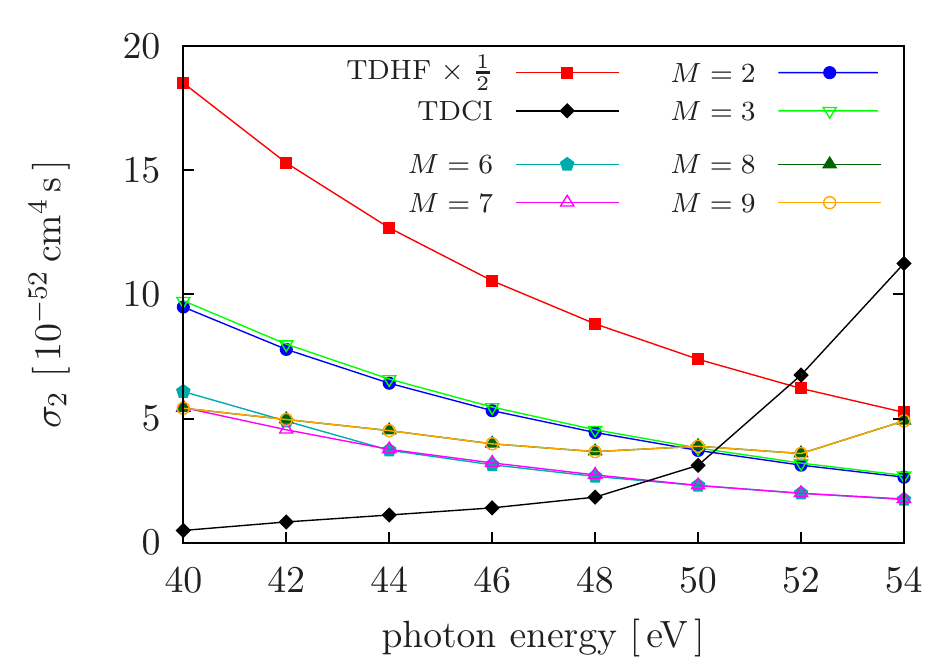}
  \end{center}
\caption{\label{fig:ion_vs_freq_acc}(Color online) Convergence behaviour of the MCTDHF double ionization cross sections. The successive approximations $M=2,3$, $M=6,7$ and $M=8,9$ give nearly identical results.}
\end{figure}

\subsection{Two-photon ionization}
In the following we present the results for the two-photon ionization induced by electromagnetic fields with frequencies in the range $40$ to $54$ eV, and intensities from $10^{13}$ to $10^{16}$ W$/$cm${}^2$. The fields are linearly polarized and have a squared-sine envelope,
\begin{align}
\mathbf E(t) \ = \ \mathbf E_0 \sin (\omega t) \sin^2 (\pi t/T)\,,
\end{align}
for $0\leq t\leq T$ and zero else. Throughout this work, we use a duration $T=100$ a.u. ($2.4$ fs), which corresponds to roughly $24$ to $32$ cycles for the applied frequencies. The single-particle basis is discretized on a radial grid that extents up to $r_\text{max}=100$ a.u., with finite elements of length $4$ a.u. containing $11$ FEDVR functions each. The maximal angular quantum numbers are $l=3$ and $m=1$. For this choice of parameters, we obtain well converged results; a change to $r_\text{max}=150$ a.u., $l=4$ and $m=2$ yielded virtually the same results. Observables are extracted at the end of the pulses. The ionization radius in Eqs.~(\ref{single_ion},\ref{double_ion}) is chosen to be $R=20$ a.u. We show results obtained in the length gauge; the velocity gauge was found to yield virtually identical results for the various test cases we considered.

In Fig.~\ref{fig:ion_vs_int} (a), we plot the single-ionization yields versus intensity for a photon energy of $45$ eV. At intensities lower than $5\cdot10^{14}\;\text{W}/\text{cm}^2$, where perturbation theory is applicable, all depicted curves are given by straight lines with a slope of one. Over the whole range of intensities, the TDCI results are well reproduced already with the first correction to TDHF, $M=2$, while $M=5$ yields full agreement.
Fig.~\ref{fig:ion_vs_int}~(b) depicts the corresponding double ionization results. They are compared to the results of Nikolopolous \emph{et al.} \cite{Nikolopoulos_2007a}, which beside a systematic shift agree well with our TDCI curve. The TDHF result shows a similar qualitative behaviour, but is more than one order of magnitude larger. The plotted MCTDHF results deviate from the expected straight lines. The $M=2$ approximation lies roughly in between TDHF and TDCI, and approaches the latter for intensities larger than $5\cdot10^{14}\;\text{W}/\text{cm}^2$. The $M=10$ result is for most of the considered intensities more accurate; however, it shows an unsatisfactory behaviour, particularly at the intensity $2\cdot10^{14}\;\text{W}/\text{cm}^2$, where a further increase of the density leads to an unphysical decrease of the ionization yield.

The largely overestimated double ionization yield obtained from TDHF is caused by the nature of the single determinant approximation. Upon insertion of the Hartree-Fock two-particle density, which is for two particles given by $d^{\text{HF}}(\mathbf r,\mathbf r')=D(\mathbf r) D(\mathbf r')/4$, in the definition (\ref{double_ion}), one obtains $P_2= P_1^{\,2}/4$. This shows that the double ionization is directly related to the single ionization, and that, as a consequence, the TDHF approximation fails to model the physical picture of double ionization correctly.
In a similar way, low-order MCTDHF approximations should be affected by this mechanism, although with increasing number of orbitals $M$, the one- and two-body density matrices and, thus, also the ionization yields become more independent from one another.

In Fig.~\ref{fig:ion_vs_freq}, we show the total cross sections for the two-photon single and double ionization obtained at a field intensity of $10^{13}\;\text{W}/\text{cm}^2$. For the used squared-sine pulses, the cross sections can be related to the ionization yields (\ref{single_ion},\ref{double_ion}) through \cite{Foumouo_2006}
\begin{align}
\sigma_1[\text{cm}^2] \ = \ 1.032 \cdot 10^{-4} \, \frac{\omega^2 P_1}{n I_0}\,,\\
\sigma_2[\text{cm}^4 \text{s}] \ = \ 2.28 \cdot 10^{-23} \, \frac{\omega^3 P_2}{n I^{\,2}_0}\,,
\end{align}
where $n$ represents the number of cycles in the pulse, $\omega$ is the frequency in eV and $I_0$ is the intensity in W/cm${}^2$. 
The single-ionization cross section in Fig.~\ref{fig:ion_vs_freq}~(a) shows a monotonic decrease with growing photon energy which is qualitatively well reproduced by all depicted curves. For $M=4$, MCDTHF shows a good agreement with the TDCI result. In Fig.~\ref{fig:ion_vs_freq}~(b) we plot the double ionization cross section. First let us focus on the TDCI results, which is compared to the results of Feist \emph{et al.} \cite{Feist_2008}, that likewise were obtained via a direct solution of the TDSE. The deviation between the curves is caused by the different methods to extract the double ionization yield (see for instance Fig. 8 in that reference, where several other methods show similar discrepancies). In \cite{Feist_2008} it has also been observed, that the expression (\ref{double_ion}) overestimates the double ionization as compared to the one obtained via projection on Coulomb waves by roughly $25 \%$ (they used an ionization radius of $R=70$ a.u.); as can be seen from our calculations, this ratio is even larger in the case of the smaller radius $R_0=20$ a.u. used here, particularly at frequencies $\omega\geq 48$ eV. We also depict the results of two experiments \cite{Hasewaga_2005,Sorokin_2007}, which are reasonably close to our TDCI result. The $M=15$ approximation fails to give a quantitatively valid description: at small photon energies, it underestimates the TDCI results, while for photon energies larger than $48$ eV the yields become too small. Further, the monotonic growth of the TDCI result is not reproduced. However, the approximation is able to qualitatively describe the increase of the curve near the threshold.

\begin{figure}[!t]
  \begin{center}
    \includegraphics[width=0.48\textwidth]{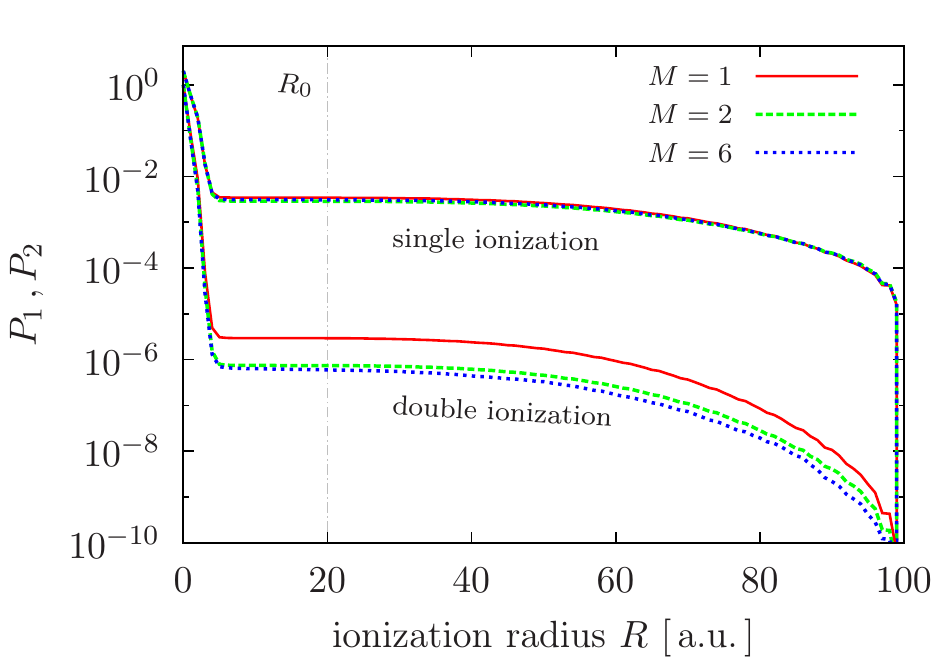}
  \end{center}
\caption{\label{fig:ion_radius}(Color online) Single and double ionization yields for selected MCTDHF approximations plotted against the radius $R_0$, at which a particle is considered ionized. The line at $R_0=20$ a.u. marks the value used in this work. The pulse parameters are $\omega=45$ eV and $I=10^{13}\;\text{W}/\text{cm}^2$.}
\end{figure}
In Fig.~\ref{fig:ion_vs_freq_acc}, we take a closer look on the convergence behaviour of the MCTDHF approximations. One can see, that TDHF again yields a far too large result for small photon energies. The approximations $M=2,3$ and $M=6,7$ improve the order of magnitude, but are not able to reproduce the increase near the threshold. The approximation $M=8$ is the first to yield a reversal of this trend around $52$ eV. It is further interesting, that the pairs of successive approximations $M=2,3$, $M=6,7$ and $M=8,9$ lead to almost identical results.

A shortcoming of our definition of ionization, Eqs.~(\ref{single_ion},\ref{double_ion}), is the arbitrariness of the ionization radius $R_0$.
In Fig.~\ref{fig:ion_radius}, we thus examine its influence on the single and double ionization yields induced by a pulse with $\omega = 45$ eV and $I=10^{13}\;\text{W}/\text{cm}^2$. One can see that for all depicted MCTDHF approximations, there is a region between $r=8$ and $r=40$ a.u., in which the yields are almost constant. From this, one can expect that for the chosen grid dimension and excitation scenario our choice of $R_0=20$ a.u. is a reasonable one. \\
\begin{figure}[!t]
  \begin{center}
    \includegraphics[width=0.48\textwidth]{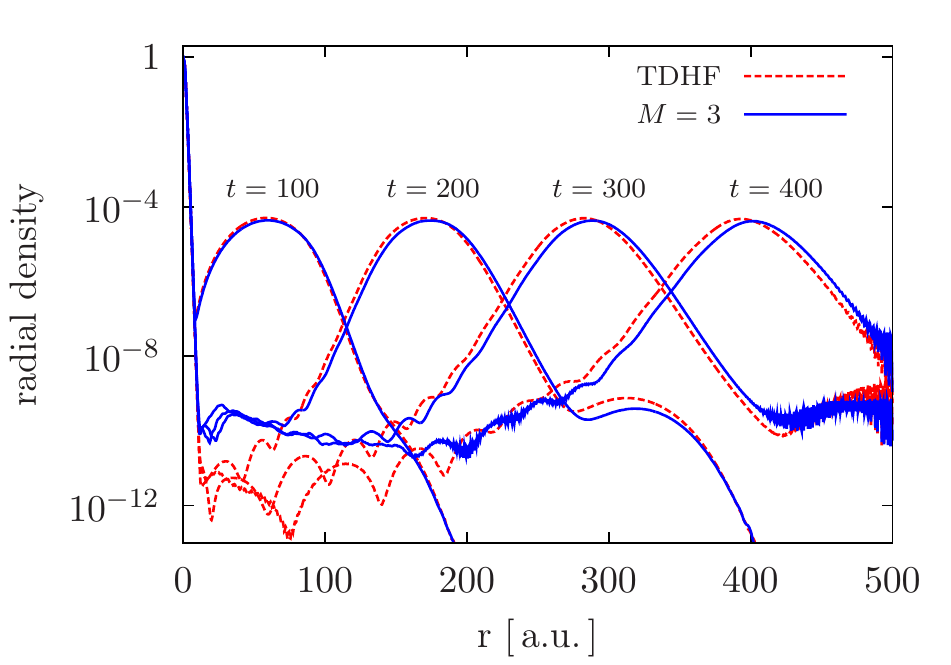}
  \end{center}
\caption{\label{fig:radial_density}(Color online) Snapshots of the radial density for different times $t$ from $100$ to $400 \; \text{a.u.}$ after application of a $42\;\text{eV}$ pulse of intensity $10^{13}\;\text{W}/\text{cm}^2$, calculated on a grid that extends up to $r_\text{max}=500$ a.u. One clearly notices a wavepacket leaving the atom, which moves slightly faster within the $M=3$ approximation than within TDHF.}
\end{figure}
The preceding results were obtained on a grid extending to $r_\text{max}=100$ a.u. in order to keep the effort of TDCI manageable. As stated before, through the use of the Poisson equation, MCTDHF provides a linear scaling with respect to the number of gridpoints. For our final result, we exploit this fact by using a five times larger grid with $r_\text{max}=500$ a.u. with the same finite element distribution as before. In Fig.~\ref{fig:radial_density}, we show snapshots of the radial density induced by a pulse with $\omega=42$ eV and $I=10^{13}\;\text{W}/\text{cm}^2$, for different times up to $t=400$ a.u. ($\sim 10$ fs). One clearly notices a wavepacket moving away from the atom; at $t=400$ a.u. a significant part of it starts to get reflected at the grid boundary, and subsequently interferes with its outgoing part. Such reflections could be  avoided relatively easily by applying exterior complex scaling or a complex absorbing potential. However, this is not necessary for the results presented in this paper since, at the time $t=100$ a.u., most of the wavepacket is still in the region inside $100$ a.u. 

One can further observe, that the velocity of the wavepacket, which is determined by difference of the photon energy and the ionization potential, is larger in the $M=3$ approximation than in TDHF. This can be explained with the well-known fact, that the Hartree-Fock approximation -- or more precisely the Koopman's theorem -- overestimates the exact value of the ionization potential. The MCTDHF approximation improves on this, and the resulting smaller ionization potential leads to a larger electron excess energy.

\section{Discussion}
The numerical investigation of the two-photon ionization of Helium presented in this paper revealed much of the capabilities and limitations of the MCTDHF formalism. Through comparison with time-dependent (full) Configuration Interaction calculations using the same single-particle basis, we were able to critically assess the accuracy of different MCTDHF approximations on the many-body level. The TDCI results itself were compared to results from the literature and showed a reasonable agreement. The conclusions drawn in the following are likely to apply also to other excitation scenarios and, more importantly, to atoms with more electrons, which is the desired area of application of MCTDHF.

First, we observed that the single-particle ionization is very well reproduced already with a small number of time-dependent orbitals ($M\sim 4$), whereas TDHF yields qualitative agreement. The effect of MCTDHF is thereby to provide a correlated background on which the ionization process takes place and which -- in contrast to single-active electron calculations -- is self-consistently found and adapted during the temporal evolution.

The double ionization is -- for the considered photon energies -- a highly correlated process, and thus a severe test for MCTDHF. Correspondingly, its correct description requires a large determinant basis. For the present case of Helium, the maximum number of $225$ Slater determinants ($M=15$) is not sufficient to achieve converged results, yet it suffices for a qualitative agreement with the fully correlated TDCI result (for which more than $6$ million Slater determinants were used).

One surprising point we observed is that the MCTDHF results become \emph{better} with increasing intensity, a behaviour totally different from approaches based on perturbation theory, such as LOPT. Accordingly, the double ionization is described better if the ionized fraction constitutes a significant part of the total particle number. On the contrary, the time-dependent variational principle performs worse in modelling a small ionized fraction, in favour of giving a well description in the vicinity of the atom. For atoms with more electrons, and for a given number $M$ of time-dependent orbitals, we however expect an improved accuracy of the double ionization as compared to the Helium case, due to the largely increased Slater determinant basis, while the difficulties encountered with low-order corrections are hopefully shifted towards higher-particle ionization yields.

The present numerical study of the MCTDHF method focused on the two-photon ionization of Helium since here the time-dependent Schr\"odinger equation is directly solvable and direct comparisons are possible. Based on these results, our future work will concentrate on the application of the MCTDHF method to the photoionization of larger atoms and molecules where direct solutions of the Schr\"odinger equation are not (yet) feasible.

\section*{Acknowledgements}
We thank Daniel J. Haxton for a useful discussion concerning the use of the Poisson equation, and Matthias Nest for valuable comments on the manuscript. This work is supported by a CPU grant at the North German Supercomputer Center (HLRN, grant shp0006), by the Bundesministerium f\"ur Bildung und Forschung (BMBF) via the project FLASH and the U.S. Department of Energy award DE-FG02-07ER54946. MB acknowledges hospitality of the KITP,  University of California Santa Barbara and partial support
by the National Science Foundation under Grant No. PHY05-51164.

\begin{appendix}
\section{Spin-restricted MCTDHF equations\label{AppA}}
The following derivation of the MCTDHF equations is mainly inspired by the work of Alon \emph{et al.} \cite{Cederbaum_unified}, who derived equations for fermions in a spin-unrestricted fashion where the spin was treated implicitly, i.e. absorbed in the orbital index. Here, we present a spin-restricted formulation. The appearance of the equations and the mathematical procedure is, apart from notational issues, essentially identical. Yet, there are differences in the interpretation: in the following spin-restricted formulation, $2M$ spin-orbitals are described by $M$ spatial orbitals, $\smallket{\phi_{p\alpha}(t)}=\smallket{\phi_{p\beta}(t)}=:\smallket{\phi_{p}(t)} $.
The indices of the matrix elements and summations correspondingly range from $1$ to $M$. The unrestricted treatment in Ref.~\cite{Cederbaum_unified} is formulated in terms of the $M_\alpha+ M_\beta$ \emph{spin}-orbitals. There, the indices run from $1$ to $M_\alpha+ M_\beta$ and roughly one half of the matrix elements are equal to zero, due to the orthogonality of $\alpha$ and $\beta$ spin-functions. This is however is not indicated in the reference, as the spin summation in the electron integrals is lacking. A further difference between the two approaches is in the appearance of the density matrices, Eqs.~(\ref{Dens1}) and (\ref{Dens2}), which are here composed of summations over the internal spin variables, whereas in Ref.~\cite{Cederbaum_unified} they are set up in terms of the elementary spin-orbital excitation operators.

Let us turn to the actual derivation, and first to the orbital equation. In the action functional (\ref{action_func}), we express the expectation value of $\hat H - i\frac{\partial}{\partial t}$ in terms of the density matrices and require the functional derivative with respect to the orbitals to vanish:
\begin{align}
\label{eqA1}\notag 0 \ &\equiv \ \frac{\delta}{ \delta \smallbra{\phi_p}} \ S\Bigl[ \bigl\{ C_{\mathbf n}  \bigr\} ,\bigl\{ \smallket{ \phi_p} \bigr\} \Bigr ] \\
&= \Biggl\{ \sum_{q} D_{pq} \; \Biggl[ \hat h \, - \, i \tderiv \ \Biggr] \; \ket{\phi_q} \, \\
\notag &+ \, \sum_{qrs} \, d_{pqrs} \; \underbrace{\braket{\phi_r}{\hat g}{\phi_s}}_{=\,\hat g_{rs}} \, \ket{\phi_q} \,\Biggr\} - \,  \sum_{q} \; \mu_{pq}(t) \; \ket{\phi_q}\,.
\end{align}
Applying the projector $\smallket{\phi_m}\smallbra{\phi_m}$, summing over $m$ and solving for the term with the Lagrange multiplier $\mu_{pm}$, we obtain:
\begin{align}
\label{eqA2}\sum_m \mu_{pm}(t) \ket{\phi_m} \, &= \, \sum_m \ket{\phi_m}\bra{\phi_m} \ \times \, \\  \Biggl\{ \sum_{q} D_{pq} \; \Biggl[ \hat h \, - \, i \tderiv \  & \Biggr] \; \ket{\phi_q} \,
\notag + \, \sum_{qrs} \, d_{pqrs} \; {\hat g_{rs}} \, \ket{\phi_q} \,\Biggr\}\,.
\end{align}
In the two Eqs.~(\ref{eqA1}) and (\ref{eqA2}), the curly brackets are equal. Insertion of the latter in the former thus yields
\begin{align}
0 \, &= \hat {\mathbf P} \; \Biggl\{ \sum_{q} D_{pq} \, \Biggl[ \hat h  -  i \tderiv \ \Biggr] \ket{\phi_q} + \sum_{qrs} d_{pqrs} \, \hat g_{rs} \, \ket{\phi_q}\Biggr\}\,,
\end{align}
where $\hat{\mathbf P}$ was defined in Eq.~(\ref{projector}).
Solving for the term with the time derivative, multiplying by $(\mathbf D^{-1})_{np}$ and summing over $p$ yields:
\begin{align}
i \, \hat {\mathbf P} \, \ket{\dot  \phi_n} \, &= \, \hat {\mathbf P} \;\Biggl\{ \hat h \, \ket{\phi_n} + \sum_{pqrs} (\mathbf D^{-1})_{np} \, d_{pqrs} \, \hat g_{rs} \, \ket{\phi_q}\Biggr\}\,.
\end{align}
In order to obtain explicit equations, we apply a unitary transformation among the orbitals after which
\begin{align}\label{unitary_trans}
\bigbraket{\phi_m}{\tderiv}{\phi_n} \, &= \, 0
\end{align}
is satisfied. This causes the projector on the lhs. to vanish and leads immediately to the orbital equation~(\ref{MCTDHF_orbital}).

For the coefficient equation we set the derivative with respect to the coefficients equal to zero. Therefore, we insert the MCTDHF expansion (\ref{MCTDHF_expansion}) into the action functional
to obtain a form which explicitly depends on the coefficients. The derivative then directly yields
\begin{align}
\notag 0 \ &\equiv \ \frac{\delta}{ \delta C_{\mathbf n}} \ S\Bigl[ \bigl\{ C_{\mathbf m}  \bigr\} ,\bigl\{ \smallket{ \phi_p} \bigr\} \Bigr ] \\
&= - i\frac{\partial C_{\mathbf n}}{\partial t} + \sum_{\mathbf m} C_{\mathbf m}\, \braket{\mathbf n}{\hat H - i \tderiv}{\mathbf m}\,.
\end{align}
With the property (\ref{unitary_trans}), also the Slater determinant matrix element of the time derivative vanishes, and we obtain the wavefunction equation.\\
Practically, the constraint (\ref{unitary_trans}) avoids rotations within the subspace spanned by the orbitals. If such a rotation is to be performed, only the wavefunction coefficients are affected, while the single-particle subspace is varied only if unavoidable. However, the applied unitary transformation is only a special representative of a broader class of constraints which read
\begin{align}
\bigbraket{\phi_m}{\tderiv}{\phi_n} \, &= \, U_{mn}(t)\,,
\end{align}
where $U_{mn}(t)$ is an arbitrary time-dependent hermitian matrix. This leads to a slightly different form of the MCTDHF equations, see Refs.~\cite{Beck_2001,Cederbaum_unified}. For instance, by choosing $U_{mn}(t) = h_{mn}(t)$ one obtains the working equations
\begin{gather}
i  \,  \dot C_{\mathbf n}(t) \ = \ \sum_{\mathbf m} \braket{\mathbf n}{ \frac 12 \sum_{pqrs} \, g_{pqrs} \, \hat e_{pqrs} }{\mathbf m} \; C_{\mathbf m}(t)\,, \\
i \,  {\ket{ \dot \phi_n}} \, = \, \hat h(t) \ket{ \phi_n} \, + \, \hat{\mathbf  P} \, \Biggl\{ \, \sum_{pqrs} \left( {\mathbf D}^{-1}\right)_{np} d_{pqrs} \, \hat g_{rs} \ket{ \phi_q} \Biggr\}\,.
\end{gather}
We tested them as an alternative for the real-time propagation and observed a roughly ten percent slower performance, in contrast to what is reported for MCTDH calculations in Ref.~\cite{Beck_2001}. This is likely to be caused by the dominant contribution of the solution of the orbital equation to the total effort.

We finally note that the previous derivation relies heavily on the orthonormality of the underlying MCTDHF orbitals, which is enforced by the Lagrange multipliers in the action functional (\ref{action_func}). If we allowed for nonorthonormal orbitals, several complications would arise, as can be seen, for instance, in the derivation of the unrestricted bosonic Hartree-Fock approximation \cite{Heimsoth_2010}. Particularly, and most intriguingly, the derivative of the density matrices with respect to the orbitals would no longer vanish.

\end{appendix}

\end{document}